# Improving EEG Signal Classification Accuracy Using Wasserstein Generative Adversarial Networks

Joshua Park, Priyanshu Mahey, Ore Adeniyi

The University of British Columbia

## Abstract

Electroencephalography (EEG) plays a vital role in recording brain activities and is integral to the development of brain-computer interface (BCI) technologies. However, the limited availability and high variability of EEG signals present substantial challenges in creating reliable BCIs. To address this issue, we propose a practical solution drawing on the latest developments in deep learning and Wasserstein Generative Adversarial Network (WGAN). The WGAN was trained on the BCI2000 dataset, consisting of around 1500 EEG recordings and 64 channels from 45 individuals. The generated EEG signals were evaluated via three classifiers yielding improved average accuracies. The quality of generated signals measured using Frechet Inception Distance (FID) yielded scores of 1.345 and 11.565 for eyes-open and closed respectively. Even without a spectral or spatial loss term, our WGAN model was able to emulate the spectral and spatial properties of the EEG training data. The WGAN-generated data mirrored the dominant alpha activity during closed-eye resting and high delta waves in the training data in its topographic map and power spectral density (PSD) plot. Our research testifies to the potential of WGANs in addressing the limited EEG data issue for BCI development by enhancing a small dataset to improve classifier generalizability.

## 1. Introduction

Electroencephalography (EEG) is a non-invasive neuroimaging technique which enables insight into brain functions, including perception, attention, memory, and motor control [1][2]. EEG's spatial and temporal resolution, alongside its non-invasive nature, make it a prime tool for use in brain-computer interfaces (BCIs), particularly for prosthetics and assistive devices [3][4]. One of the major challenges for BCIs, however, is the limited quantity of EEG data [5][6][7][8]. The variance of EEG signals across individuals and tasks also poses an issue [9]. Additionally, EEG data has a small signal-to-noise ratio, making it challenging to extract relevant information from the recorded signals [10]. To address limited EEG datasets and high variability that lead to difficulty in the development of EEG-based BCI, Generative Adversarial Networks (GANs) have emerged as a solution.

GANs consist of two neural networks: the generator and the critic [11]. The generator and critic are engaged in a zero sum game where the generator attempts to generate synthetic signals and the critic is tasked with discriminating between that synthetic data and real data. One of the key advantages of GANs is their ability to understand the distribution of feature range and frequency in training data which allows

GANs to generate new and similar—but not identical—instances of data [11]. GANs have already been proven to be capable of creating realistic synthetic data for real-world datasets [12].

Traditional GANs have limitations such as mode collapse and vanishing gradients issues [13], which negatively impact both the generator and critic performance. To overcome traditional limitations, WGANs provide a promising alternative; WGANs, unlike traditional GANs, use the Earth-Mover (Wasserstein-1) distance as the optimization objective with the goal to minimize the dissimilarity of the generated data distribution to the real data distribution [13]. Additionally, WGANs typically pair with a gradient penalty term that imposes a Lipschitz constraint on the critic's weights. This prevents unnecessary learning fluctuations during the optimization step, leading to stable training, more robust gradients, and mitigation of the vanishing gradient problem [13]. The gradient penalty in WGAN is applied by integrating a mixture of real and generated signals into the critic's loss function during each training iteration, facilitating smoother transition and continual improvement over time [14][15]. Given enhanced stability and robustness, this paper will utilize WGANs to generate realistic EEG signals.

## 2. Methods

### 2.1 Dataset

The WGAN model was trained on the BCI2000 dataset [16] which was stored in EDF format with 64 electrodes in 10-10 format (excluding electrodes Nz, F9, F10, FT9, FT10, A1, A2, TP9, TP10, P9, and P10). The dataset consists of about 1500 one-minute and two-minute EEG recordings from 109 people performing various motor-imagery tasks. Our model utilized eyes-open and eyes-closed data. Each chosen sample was 61 seconds long with a sampling rate of 160 with 64 channels per sample. 45 samples out of 109 were chosen after filtering for incomplete or unusually noisy data. Originally, samples were of 9760 sequence length (61s x 160 sampling rate). Given the repetitive nature of the data, we truncated the sequence length to 3152. Additionally, the dataset was normalized to an interval of [-1, 1]. No further preprocessing was done.

### 2.2 Model Training

Our proposed model's loss function applies Wasserstein loss to approximate and minimize the Earth Mover's Distance (EMD) between the real and fake feature distributions. By using the two distributions' distance as the loss score, the critic does not need to output a score bounded between [-1,1] which helps alleviate the vanishing gradient and mode collapse problem [13]. The gradient penalty is an important regularization term that enforces the critic to be 1-Lipschitz continuous [15]. By constraining the loss function to be Lipschitz, it enforces continuity and differentiability, achieving more stable results [15]. Together, our loss function is represented as follows:

$$L = E_{x \sim p_g}[C(G(z))] - E_{x \sim p_r}[C(x)] + \lambda E_{x \sim p_{\hat{x}}}[(||\nabla C(\hat{x})||_2 - 1)^2]$$

$$\min_G \max_C E_{x \sim p_g}[C(G(z))] - E_{x \sim p_r}[C(x)]$$

The critic $C$ aims to maximize the average cost ($E(C(x))$) for the distribution of real values $r$, and the generator $G$ aims to minimize the average cost ($E(G(z))$) for the distribution of fake values $g$. As the critic learns to evaluate real and fake samples, it should return smaller Wasserstein distance for real samples $x$ and large Wassertein distance for generated samples $\tilde{x}$. By subtracting the two loss scores, the

WGAN's ultimate goal is to minimize loss function $L$. The gradient penalty term penalizes the loss function when the critic's gradient norm is greater than 1 [15], which is how it enforces the Lipschitz constraint [15]. To calculate the regularization term, we used interpolated samples $\hat{x}$ which is an intermediate sample between real and fake samples. Intermediate samples are represented by $\hat{x} = \epsilon x + (1 - \epsilon)G(z)$ where $\epsilon$ is a random number used to pick a random point in space between real and fake data [15].

Our WGAN model's training algorithm follows a similar approach to the conventional WGAN training [15]. The noise vector $z$ is first sampled from a Gaussian distribution and fed into the generator, which is then transformed into fake samples closely resembling the real data ($\tilde{x} = G(z)$). The fake sample is evaluated by the critic and the generator uses the critic's feedback to output more realistic fake samples. We repeat these steps until the generator's output is indistinguishable from real data. The objective is to eventually trick the critic into believing the generator's output is real data.

### 2.3 Model Architecture

Our proposed model was trained for 50,000 epochs with a batch size of 32. Weights were initialized with Gaussian distribution of mean 0 and standard deviation of 0.02. Similarly, noise vectors were generated with a latent size of 500, each value being drawn from the Gaussian distribution of mean 0 and standard deviation of 0.02. Since critics typically learn faster than generators [13], the critic was trained 5 times for every training iteration of the generator to maintain a consistent learning rate between the two neural networks. The gradient penalty was set to 10 and batch normalization was only used for the generator. We used LeakyRELU with an alpha of 0.2 as our activation function over a classic RELU, due to its advantage in avoiding vanishing neurons [17]. The architecture of the neural networks can be found in Table S1 with both networks consisting of hidden channels = 150 and kernel size = 3, stride = 1, and padding = 0. We used Adam optimizer with a learning rate of 0.0001 and betas = (0.5, 0.99). For upsampling and downsampling, we used linear interpolation and average pooling respectively to minimize unwanted frequency artifacts [18].

### 2.4 Classification Performance

The generated EEG signals were evaluated based on the performance of three classifiers: a fully connected neural network (FNN), a convolutional neural network (CNN), and a recurrent neural network (RNN). By comparing the accuracies of classifiers trained on the real data to classifiers trained on a combination of real and generated data, we can assess the generated EEG signals' effectiveness in augmenting the available data in a practical setting [19]. To train our classifiers, we used an 80/20 split for the train and test datasets.

## 3. Results

### 3.1 Power Spectral Density

To evaluate the frequency properties of the real and generated EEG signals, we plotted the PSD using Welch's averaged periodogram method with a Hanning window of length 256 with a 50% overlap. The PSD plots for eyes-closed and eyes-open are shown in Figure 1. They are the averaged result over all of the samples and 64 channels.

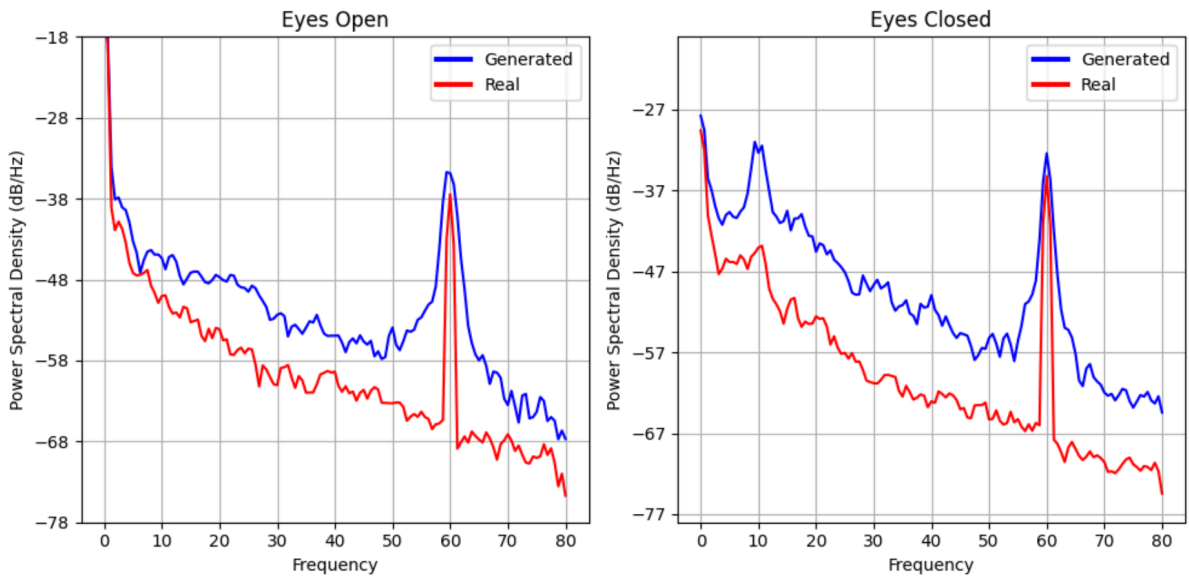

Figure 1. PSD plot of real and generated signals for eyes-open and eyes-closed

### 3.2 Classification Accuracy and FID Score

We trained each of our CNNs, FNNs, and RNNs with both real data and a mix of real and generated data, using identical hyperparameters and loss functions. The following accuracies are averages from 100 trials and showcase significant improvements for each model (Table 1). To validate the reliability of our generated data, we computed FID scores averaged over 100 trials (Table 2). We calculated the FID score using the features extracted by our trained CNN model.

|  | CNN | RNN | FNN |
|---|---|---|---|
| Real data | 79.296% | 75.835% | 64.037% |
| Real data and generated data | 96.526% | 87.918% | 83.829% |
| **Improvement** | **+17.23%** | **+12.083** | **+19.792** |

Table 1. Average classification accuracy based on different classifiers

|  | Eyes-open | Eyes-closed |
|---|---|---|
| Real data and real data | 0.0 | 0.0 |
| Real data and noise | 4866.5983 | 1083.0996 |
| Real data and generated data | **1.345** | **11.565** |

Table 2. Average FID scores for different data types of eyes-open and eyes-closed

### 3.3 Topographic Mapping

The topographic maps of our 64-channel EEG data were plotted using the mne-python library. Sensor locations were visually marked, while sensor labels were disregarded. The EEG data was enhanced with contour lines drawn for amplitude distribution, and outlines representing a standard head model were incorporated (Figure 2 and Figure 3).

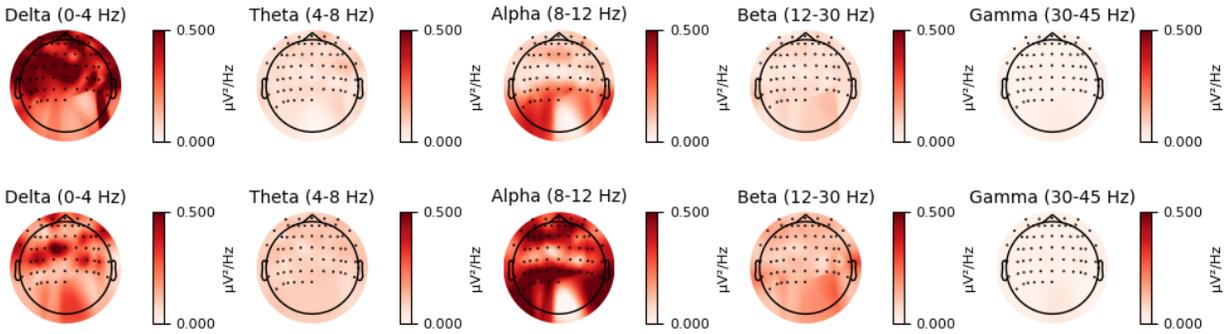

Figure 2. Normalized topographic map of eyes-closed comparing real (top) and generated (bottom) signals

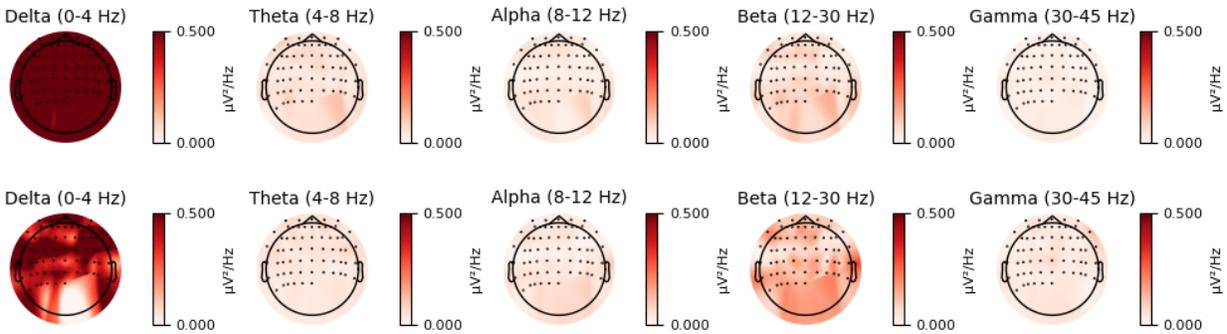

Figure 3. Normalized topographic map of eyes-open comparing real (top) and generated (bottom) signals

## 4. Discussion

The data generated by the WGAN proved to be representative of real EEG data, as indicated by the FID score and the topographic map. The FID score uses a trained classifier and has been shown to be equivalent to human judgment [20]. Previous research has shown that there is no direct correlation between the FID score and the WGANs ability to represent the spectral and spatial properties of EEG [21]. We do note however, that such studies were based on an external model while our model was trained in-house. The FID score was consistently worse for eyes-closed, regardless of the classification model we used. Furthermore, the WGAN had an architecture complex enough to mimic the spectral properties of the time-series EEG training data without adding any spectral loss term to the loss function like some WGAN models choose [22]. The PSD plots lack fidelity but they do, however, mimic the shape of real EEG. Similarly, the generated data's topographic map has denser signals compared to the real data but generally resembles its density and signal distribution.

EEG signals can be broken down into different frequency bands (delta, theta, alpha, beta, and gamma), each of which are associated with different cognitive and physiological states. The large alpha wave oscillation is caused by blocking of visual inputs to the occipital region when eyes are closed [23]. Our generated data reflects this physiology with increased alpha wave activity during the eyes-closed stimulus. Our model observes high delta waves and this was most likely due to the high presence of the delta waves seen in the training data [24]. In some cases, other papers opted to first switch to image-based neuroimaging techniques and used that as the GANs' training data [25][26]. In our study, the WGAN was able to mimic the spectral and spatial nature of alpha signals while eyes were closed, as shown by the PSD and topographic map. Our model was able to pick up on key differentiating features, in this case being the larger alpha waves in eyes-closed stimulus.

Since WGANs extrapolate and generalize the feature distribution of the training data, the training dataset for the classifier becomes denser and more diverse. The generated samples improve the generalizability of the classifier by providing a generalized view of data variations [27]. By expanding a small dataset, the classifier performs better on unseen data and reduces overfitting [28]. As shown by Table 1, all classifiers performed better with the WGAN-enriched dataset.

### 4.1 Limitations

Our investigation relied on just one dataset for training the WGAN. We cannot conclusively affirm the model's generalizability across multiple or different datasets; its effectiveness may be rather confined to the specific task of distinguishing between eyes-open and eyes-closed scenarios and its efficiency on other EEG signal classifications, such as emotion classification, remains unexplored.

Additionally, we trained two separate WGANs for eyes-open and eyes-closed scenarios, which limited our capacity to pair a person's unique EEG characteristics for these two states. Every individual can potentially exhibit unique characteristics in their EEG signals, which would ideally be reflected as a pair of eyes-open and eyes-closed states. However, our model does not capture these individual specific nuances, potentially causing mismatch in analyses and interpretations of the generated data.

### 4.2 Future work

Future research should focus on practical BCI applications to evaluate the actual impact of EEG classification accuracy improvement. The current study demonstrated an increase in classification accuracy of deep learning classifier models by integrating real and artificially generated EEG data. However, we suggest conducting similar experiments in real-world BCI applications, such as neuroprosthetics, neurofeedback, and other assistive technologies. Evaluations should be performed in real-time conditions to understand how generated EEG signals affect the performance of BCI applications. Further explorations can also be carried out on identifying optimal methods to combine real and generated EEG data for different types of BCIs to improve our method's adaptability. Additionally, expanding the experiment to include a broader range of EEG classes (rather than just eyes-open and closed states) and utilizing more complex deep learning techniques could also strengthen future research endeavors.

## 5. Conclusion

Our study supports the potential of using WGANs with gradient penalty to enrich EEG datasets and address the issue of limited training data in BCIs. We observed increased classification accuracy in CNN, RNN, and FNN classifiers when trained with both real and WGAN-generated data. Further, low FID scores and resemblance of power spectral density patterns and topographic mappings between real and generated data indicated the realistic representation of the latter. The realism and data variety created by WGANs suggest that WGANs can significantly enhance BCI performance and efficiency by expanding and diversifying EEG datasets.

# Supplementary

| Generator | | | Critic | | |
|---|---|---|---|---|---|
| Layer | k,s,p | Output shape | Layer | k,s,p | Output shape |
| Input (z), Dense | | (8100) | Input | | (64, 3152) |
| Conv, BatchNorm, LRELU | 3, 1, 0 | (150, 52) | Conv, LRELU | 3, 1, 0 | (150, 3150) |
| Upsample (linear interpolation) | | (150, 106) | Conv, LRELU | 3, 1, 0 | (150, 3148) |
| Conv, BatchNorm, LRELU | 3, 1, 0 | (150, 104) | Downsample (average pooling) | | (150, 1576) |
| Conv, BatchNorm, LRELU | 3, 1, 0 | (150, 102) | Conv, LRELU | 3, 1, 0 | (150, 1574) |
| Upsample | | (150, 206) | Conv, LRELU | 3, 1, 0 | (150, 1572) |
| Conv, BatchNorm, LRELU | 3, 1, 0 | (150, 204) | Downsample | | (150, 787) |
| Conv, BatchNorm, LRELU | 3, 1, 0 | (150, 202) | Conv, LRELU | 3, 1, 0 | (150, 785) |
| Upsample | | (150, 406) | Conv, LRELU | 3, 1, 0 | (150, 783) |
| Conv, BatchNorm, LRELU | 3, 1, 0 | (150, 404) | Downsample | | (150, 392) |
| Conv, BatchNorm, LRELU | 3, 1, 0 | (150, 402) | Conv, LRELU | 3, 1, 0 | (150, 390) |
| Upsample | | (150, 806) | Conv, LRELU | 3, 1, 0 | (150, 388) |
| Conv, BatchNorm, LRELU | 3, 1, 0 | (150, 804) | Downsample | | (150, 195) |
| Conv, BatchNorm, LRELU | 3, 1, 0 | (150, 802) | Conv, LRELU | 3, 1, 0 | (150, 193) |
| Upsample | | (150, 1606) | Conv, LRELU | 3, 1, 0 | (150, 191) |
| Conv, BatchNorm, LRELU | 3, 1, 0 | (150, 1604) | Downsample | | (150, 96) |
| Conv, BatchNorm, LRELU | 3, 1, 0 | (150, 1602) | Conv, LRELU | 3, 1, 0 | (150, 94) |
| Upsample | | (150, 3206) | Conv, LRELU | 3, 1, 0 | (150, 92) |
| Conv, BatchNorm, LRELU | 3, 1, 0 | (150, 3204) | Downsample | | (150, 45) |
| Conv | 1, 1, 0 | (64, 3204) | Conv | 1, 1, 0 | (64, 45) |
| Dense | | (64, 3152) | Dense | | (64, 1) |

Table S1: WGAN critic and generator model architecture

The python script used for data preparation and modeling is stored in this public repository:
https://github.com/JoshParkSJ/eeg-WGAN

The dataset we used for the WGAN model is stored here: https://physionet.org/content/eegmmidb/1.0.0/
[16]